\documentclass[aps,prl,twocolumn,superscriptaddress]{revtex4}

\usepackage{graphicx}

\pdfoutput=1

\begin{document}

\title{Quantum dot spectroscopy using cavity QED}

\author{Martin Winger}
\author{Antonio Badolato}
\author{Kevin J.\ Hennessy}
\affiliation{Insitute of Quantum Electronics, ETH Zurich, 8093 Zurich, Switzerland}
\author{Evelyn L.\ Hu}
\affiliation{California NanoSystems Institute, Universtiy of California, Santa Barbara, California 93106, USA}
\author{Ata\c c Imamo\u glu}
\affiliation{Insitute of Quantum Electronics, ETH Zurich, 8093 Zurich, Switzerland}

\date{\today}

\begin{abstract}
Cavity quantum electrodynamics has attracted substantial interest,  both due to its potential role in
the field of quantum information processing and as a testbed for basic experiments in quantum
mechanics. Here, we show how cavity quantum electrodynamics using a tunable  photonic crystal
nanocavity in the strong coupling regime can be used for single quantum dot spectroscopy. From the
distinctive avoided crossings observed in the strongly coupled system we can identify the neutral and
single positively charged exciton as well as the biexciton transitions. Moreover we are able to
investigate the fine structure of those transitions and to identify a novel cavity mediated mixing of
bright and dark exciton states, where the hyperfine interactions with lattice nuclei presumably play
a key role. These results are enabled by a deterministic coupling scheme which allowed us to achieve
unprecedented coupling strengths in excess of $150\ \mathrm{\mu eV}$.
\end{abstract}

\pacs{}

\maketitle

Cavity quantum electrodynamics (QED) studies the quantum limit of light-matter interaction in an
optical cavity \cite{Mabuchi:Science}; its implementation in both atomic \cite{Raimond:RMP} and solid
state systems \cite{Michler:Science} has received substantial interest primarily due to potential
applications in quantum information processing \cite{Mabuchi:Science,Imamoglu:PRL}. Self-assembled
quantum dots (QDs) embedded in monolithic nanocavities provide a well-controlled and robust system
for conducting cavity QED experiments in the solid state. The combination of ultra-small mode volumes
and high Q values provided by those systems has made it possible to enter the strong coupling regime
of cavity QED with different types of monolithic cavities
\cite{Yoshie:Nature,Peter:PRL,Reithmaier:Nature,Hennessy:Nature}. However, in all previously reported
experiments, a nanocavity mode was brought into resonance with a QD transition whose nature could not
have been specified. This is mainly due to the fact that the inherently random excitation scheme used
in photoluminescence (PL) spectroscopy allows for the creation of a multitude of QD charge
configurations and the resulting spectral emission lines are generally difficult to interpret. This
ambiguity has not hindered progress in the field since the emphasis has so far been on using the QD
as a two-level emitter that couples resonantly to a nanocavity mode
\cite{Vuckovic:Nature,Painter:Nature}; the nature of the particular QD transition was irrelevant for
the study of cavity-QED physics.

In contrast, we show here that a nanocavity mode that is strongly coupled to QD transitions can be
used as a powerful spectroscopic tool for studying the fundamental properties of the QD itself. The
distinct spectral anticrossings observed when the cavity is tuned across resonance with different QD
transitions allow us first of all to unambiguously identify the single (positively) charged exciton
($X^+$), the neutral exciton ($X^0$), and the neutral biexciton ($XX^0$) emission lines.
Surprisingly, cavity-QED also allows us to study the fine structure of QD transitions
and identify a novel strong coupling induced mixing of bright and dark exciton complexes.

In order to reach the strong coupling  regime of cavity QED the coherent coupling rate $g$ between an
excitonic transition and a cavity field has to exceed the decay rates of the exciton ($\Gamma$) and
the cavity ($\kappa$), obeying $g>\Gamma/4,\kappa/4$ \cite{Andreani:PRB}. This is achieved in a
deterministic fashion using the approach introduced in \cite{Hennessy:Nature,Badolato:Science} where
cavities are precisely positioned around pre-selected QDs in order to achieve a maximum coupling
strength. InAs QDs were grown by molecular beam epitaxy in the centre of a $\mathrm{126\ nm}$ thick
GaAs membrane on top of a $1\ \mathrm{\mu m}$ thick $\mathrm{Al_{0.7}Ga_{0.3}As}$ layer that allowed
for subsequent cavity membrane formation. In situ annealing shifted the QD s-shell emission
wavelength to $\lambda \simeq 940\ \mathrm{nm}$. We identified well-isolated QDs in PL and mapped
them using atomic force microscopy. Around those QDs we then fabricated PC defect cavities in the L3
configuration \cite{Akahane:Nature} using electron beam lithography in combination with a dry and wet
etching process. The effective mode volume of an L3 cavity is $V_{\mathrm{eff}} = 0.7 \cdot
(\lambda/n)^3$ and the theoretical $Q$-value is $Q = 3 \cdot 10^5$. In general, the positioning
technique we use has a precision of approximately 30 nm. The device we study here has negligible
misalignment between the QD and the field maximum of the first order cavity mode, such that we expect
a maximum coherent coupling rate $g$.

In order to perform micro-PL  studies at $T = 4.2\ \mathrm{K}$, the sample was mounted into a liquid
Helium flow cryostat. For off-resonant excitation of the GaAs host material we used a titanium
sapphire laser that could be either operated in continuous wave mode or mode-locked in order to emit
picosecond pulses. The excitation laser beam was focused onto the PC cavity using a microscope
objective with a numerical aperture of $N\! A = 0.55$. The luminescence signal was collected through
the same microscope objective, coupled into a single mode fibre, and directed to a grating
spectrometer for spectral analysis. A holographic grating with 1500 groves per millimetre allowed for
a spectral resolution of $30\ \mathrm{\mu eV}$. The signal was detected with a liquid nitrogen cooled
charge coupled device (CCD) camera. Moreover, a crystal polarizer in the detection path allowed for
doing polarization sensitive PL studies.

\begin{figure}
    \vspace*{.05in}
    \includegraphics{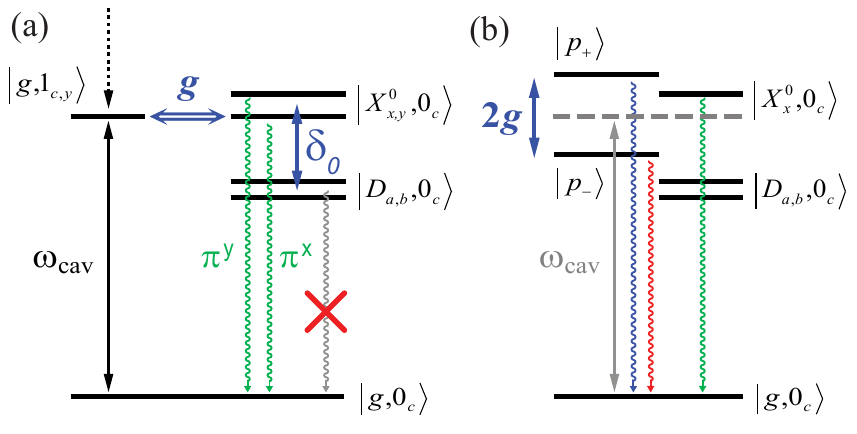}
    \caption{\label{Fig1} $X^0$ level scheme. (a) Eigenstates of the QD and the one-photon state of the cavity mode. The electric field vector of the cavity field is parallel to the y-direction at the QD location, giving rise to coherent coupling $g$ with $|X_y^0,0_c\rangle$ while $|X_x^0,0_c\rangle$ is not influenced by the cavity mode. (b) Dressed eigenstates of the coupled cavity-QD system. The polariton states $|p_+\rangle$ and $|p_-\rangle$ that are split by $2g$.}
\end{figure}

In the lowest energy conduction band states of a QD, electrons have spin z-component $S_{e,z} = \pm
\frac{1}{2}$, while the (heavy) holes in the lowest energy valence-band states carry pseudospin
$J_{h,z} = \pm \frac{3}{2}$. The light hole states with $J_{h,z} = \pm \frac{1}{2}$ are split off by
$> 35\ \mathrm{meV}$ and can therefore be neglected. The total angular momentum projection of the
lowest energy electron-hole pairs (excitons) therefore can take the values $M = \pm1,\pm2$. The
states with $M = \pm2$ (dark excitons) can not recombine optically, since their decay requires an
angular momentum transfer of $2\hbar$. The states with $M = \pm1$ on the other hand, give rise to the
$X^0$ luminescence. The electron-hole exchange interaction has the effect of (i) splitting dark and
bright states by $\delta_0$ on the order of $100-200\ \mathrm{\mu eV}$ and (ii) lifting the
degeneracy of both bright and dark manifolds \cite{Bayer:PRB,Ivchenko:PSS,Bester:PRB}. The coupled
excitonic eigenstates then are $|D_{a,b}\rangle = \left(|+2\rangle\pm|-2\rangle\right) / \sqrt{2}$
for the dark exciton doublet and $|X_{x,y}^0\rangle = \left(|+1\rangle\pm|-1\rangle\right) /
\sqrt{2}$ for the bright doublet. The latter decay by emitting photons that are linear polarized
along either of the two QD axes ($x,y$). This level scheme is shown in figure~\ref{Fig1}(a) along
with the lowest excited state $|g,1_{c,y}\rangle$ of the cavity-mode containing a single photon.

The electric field of the cavity mode in the centre of the PC defect, i.e.\ at the  location of the
QD, is oriented along the y-direction (perpendicular to the L3 defect line). Therefore we assume
maximum coupling $g$ between the cavity single-photon state $|g,1_{c,y}\rangle$ and
$|X_y^0,0_c\rangle$; the second bright exciton state $|X_x^0,0_c\rangle$  is not coupled to the
cavity. This is schematized in figure~\ref{Fig1}(b) where the dressed QD-cavity states are shown for
the cavity on resonance with $|X_y^0,0_c\rangle$. While the uncoupled state $|X_x^0,0_c\rangle$
remains unchanged, $|X_y^0,0_c\rangle$ hybridizes with $|g,1_{c,y}\rangle$, thereby forming the polariton
states $|p_\pm\rangle = ( |X_y^0,0_c\rangle \pm |g,1_{c,y}\rangle ) / \sqrt{2}$.

\begin{figure}
    \vspace*{.05in}
    \includegraphics{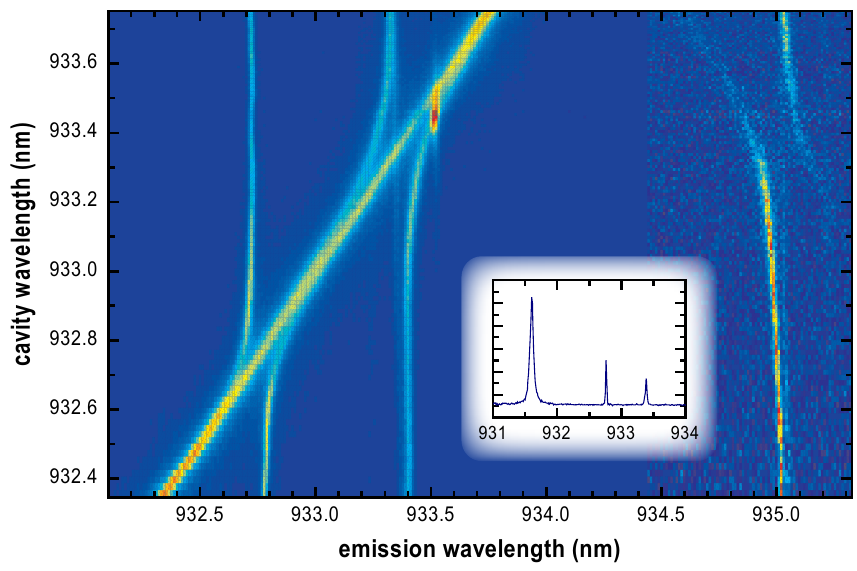}
    \caption{\label{Fig2} Colour plot of PL spectra when tuning the cavity mode across the $X^+$ and the $X^0$ transitions. For both lines a clear anticrossing can be observed. The inset shows a typical PL spectrum for the cavity blue detuned from the $X^+$ line. The excitation conditions are $P_{\mathrm{ex}} = 10$ nW and $\lambda_{\mathrm{ex}} = 818$ nm. For wavelengths on the red of 934.5 nm the colour scale has been offset by a factor of 26 in order to highlight the biexciton emission.}
\end{figure}

The inset in figure~\ref{Fig2}(a) shows a typical PL spectrum  obtained from a QD for a pump
wavelength of $\lambda_{\mathrm{ex}} = 818\ \mathrm{nm}$ and a pump power of $P_\mathrm{{ex}} = 10\
\mathrm{nW}$.
We assign the line denoted with $X^0$ to spontaneous recombination of the states
$|X_x^0,0_c\rangle$ and $|X_y^0,0_c\rangle$, which are typically split by $0-30\ \mathrm{\mu eV}$. In
the particular QD we study here, this splitting lies below the resolution of our spectral apparatus.
The identification of this line as $X^0$ becomes obvious in the results presented below. Moreover we
tentatively assign the line denoted by $X^+$ to emission from the positively charged exciton, where
in addition to the electron-hole pair an extra hole is trapped within the QD. The splitting of $0.5\
\mathrm{nm}$ from the $X^0$ is in agreement with previous studies of similar QDs. We remark here that
in order to obtain the PL spectrum depicted in the inset of figure~\ref{Fig2}(a), we had to carefully
study the pump-wavelength dependence of the PL: we find that the strong cavity-feeding effect
reported earlier \cite{Hennessy:Nature} only allow the observation of strong QD transitions for a
limited range of pump laser wavelengths.

The peak around $931.5\ \mathrm{nm}$ corresponds to  emission from the uncoupled cavity mode. This
feature can be perfectly fitted with a Lorentzian curve with a $Q$-factor of 16,000 which corresponds
to a decay rate of $\kappa = 83\ \mathrm{\mu eV}$. As has been shown in our previous work, efficient
emission from the cavity mode is always present, irrespective of the detuning from QD spectral
features \cite{Hennessy:Nature}. Photon cross-correlation measurements revealed that this emission is
solely due to the presence of a single QD in the PC membrane. The light emitted by the cavity is
predominantly polarized along the y-direction (degree of polarization = 96\%), i.e.\ perpendicular to
the L3 defect direction.

In order to tune the cavity mode into resonance with the QD excitons, we  employed a thin film
gas-deposition technique to continuously red-shift the cavity mode wavelength
\cite{Strauf:APL,Srinivasan:APL}. This technique allowed for a maximum tuning range of approximately
$6\ \mathrm{nm}$, sufficient to bring the cavity mode on resonance with both the $X^+$ and the $X^0$
lines of the QD. Figure~\ref{Fig2} shows PL data for different cavity mode wavelengths in a colour
plot. For each spectrum recorded, the central wavelength of the (uncoupled) cavity mode has been
extracted and used to linearise the vertical axis. Furthermore, since the total intensity of the
spectra fluctuates in time as a result of sample drift, each spectrum has been normalized to its
integral. In the data shown here, the polarizer was oriented along the y-direction in order to
maximize the signal from the cavity mode. As we cross the shorter wavelength line ($X^+$), we observe
an anticrossing of the cavity and QD lines along with a central peak that corresponds to emission
from the uncoupled cavity mode at times when the QD occupies a charging state other than the $X^+$.
The vacuum Rabi splitting on the $X^+$ line is found to be $2g = 205\ \mathrm{\mu eV}$. The strong
coupling condition $g > \kappa/4,\Gamma/4$ is obviously well fulfilled for those parameter values.
Moreover, we observe that the emission of the strongly coupled polariton doublet is co-polarized with
the cavity mode.

As gas tuning proceeds, the cavity moves in resonance with the $X^0$ peak in the spectrum.  The two
anticrossing lines again correspond to emission from the polariton states $|p_\pm\rangle$ that here
show a vacuum Rabi splitting of $2g = 316\ \mathrm{\mu eV}$, which is, to the best of our knowledge,
the largest splitting reported so far in any cavity-QED system with a single emitter. The
approximately 1.5 times larger coupling to this peak compared to the one observed with the $X^+$ line
is consistent with our identification of this peak with the $X^0$ line and arises from the fact that
the $X^+$ trion transitions are circularly polarized.

Furthermore, we observe that the neutral biexciton line $XX^0$ at $\lambda_{XX^0} = 935\ \mathrm{nm}$
undergoes a splitting as the cavity is on resonance with the $X^0$. This can clearly be seen as the
anticrossing feature around that wavelength in figure~\ref{Fig2}. Since the pump power here is well
below the saturation of the QD, emission from the $XX^0$ is rather weak. For this reason the colour
scale of figure~\ref{Fig2} has been offset for wavelengths longer than $934.5\ \mathrm{nm}$. Since
the $XX^0$ emission occurs via decay of the biexciton state to the neutral exciton
$|X^0_{x,y}\rangle$ states, the vacuum Rabi splitting of the $|X^0_y\rangle$ state leads to a
splitting in the $XX^0$ line. The upper (lower) polariton state shifts the biexciton decay to lower
(higher) emission energy: therefore, the anticrossing feature of $XX^0$ appears horizontally flipped
with respect to that of the $X^0$. Since the PL data shown here is obtained using a polarizer that
was oriented along the y-direction, the unsplit $XX^0$ emission to the $|X^0_{x}\rangle$ state is not
observable.

The splitting of the two anticrossing $XX^0$ branches amounts to $0.22\ \mathrm{nm}$,  identical to
the vacuum Rabi splitting of the $X^0$ itself. Moreover, the intensity of the cavity-like branches
decays when moving away from resonance. This is exactly the behaviour expected from the $XX^0$
decaying to the QD-like component of the intermediate polariton states $|p_\pm\rangle$. Furthermore,
we note that there is no signature of an uncoupled cavity peak as in the anticrossing features
observed on the $X^+$ and $X^0$ lines. This supports the hypothesis that its appearance in the latter
is not an intrinsic feature of the resonantly coupled QD-cavity system itself, but rather a
manifestation of off-resonant cavity feeding at times when the QD occupies a charging state other
than the one the cavity is resonantly coupled to.

\begin{figure}
    \vspace*{.05in}
    \includegraphics{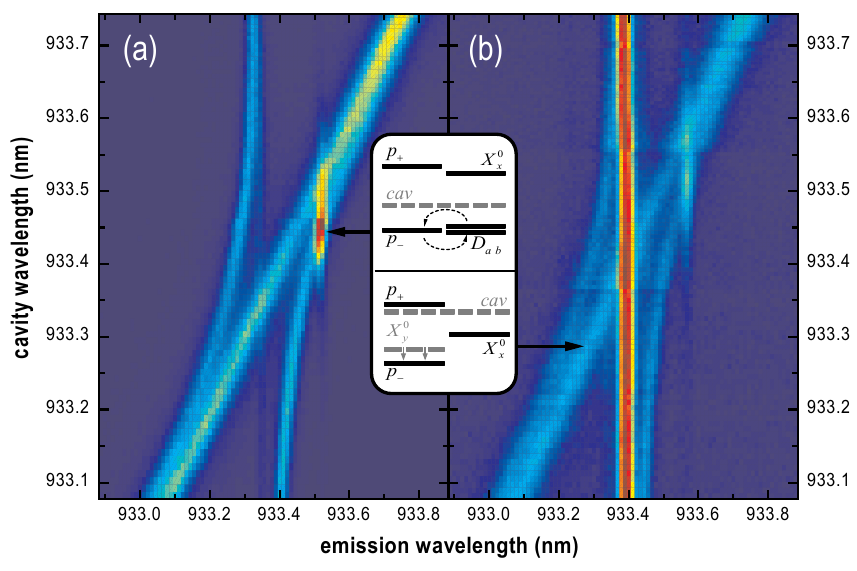}
    \caption{\label{Fig3} Comparison of $X^0$ dynamics for different emission polarization
    (a) Zoomed-in version of figure~\ref{Fig2} for the polarizer set in order to maximize cavity emission. The inset shows level schemes for two different positions of the cavity mode. In the lower inset the cavity is red-detuned from the $X^0$. As the cavity photon energy is reduced, $|p_-\rangle$ moves in resonance with the dark states $|D_{a,b}\rangle$ as indicated in the upper inset.
    (b) Here the polarizer is set to $55^\circ$ with respect to the cavity emission. Here the PL is plotted in a logarithmic colour scale.}
\end{figure}

In comparison to the anticrossing observed on the $X^+$ line, the  resonant signature on $X^0$ shows
additional features. First we notice that there is a weak additional emission line that does not
shift as the cavity is tuned across resonance. This emission can be attributed to the uncoupled state
$|X_x^0,0_c\rangle$, the dipole of which is orthogonal to the cavity electric field such that it does
not interact with the cavity mode. Further evidence comes from the polarization behaviour of the
emission: As the polarizer in the detection path is rotated, the emission from this central line
increases and dominates the spectrum for a polarization setting orthogonal to the cavity mode
emission. Figure~\ref{Fig3} shows a zoom-in of the $X^0$ strong coupling feature for two different
polarizations.
In figure~\ref{Fig3}(a) the polarizer is oriented parallel to the cavity mode emission
direction, whereas in \ref{Fig3}(b) it is rotated by $55^\circ$, in order to make emission from both
the uncoupled exciton and the polariton features visible in the same spectrum.

A corresponding fine structure can be observed on the biexciton line,  in that the anticrossing
induced by vacuum Rabi splitting of the $X^0$ is only observed for a polarization parallel to the
cavity mode (not shown in Fig.~\ref{Fig3}). The other pathway for biexciton decay leads through the
intermediate state $|X^0_x\rangle$ which does not couple to the cavity. Therefore we also observe an
x-polarized uncoupled biexciton line that does not shift as the cavity mode is being tuned. The
presence of an uncoupled exciton and biexciton lines for x-polarized PL along with the correlated
anticrossings of the y-polarized $X^0$ and the $XX^0$ emission lines make the identification of the
neutral and biexciton transitions unambiguous.

We note that the behaviour of the $X^0$ line differs significantly  from that of the $X^+$ line. In
particular, for the $X^+$ transition there is no uncoupled exciton, as is expected for a charged
exciton state. Due to the zero total spin-projection of the holes in the initial state of the $X^+$
decay, no exchange splitting occurs for the positively charged exciton. The two possible electron
spin configurations then lead to degenerate circularly polarized transitions, either of which couple
with equal strength to the cavity field. As decay then proceeds mainly via the cavity mode, the
emission is co-polarized with the cavity mode for both initial spin configurations. Experimentally,
we confirmed that there is no unsplit $X^+$ line by observing identical anticrossing features in both
polarization channels.

As tuning proceeds further to the red side of the $X^0$ line, an unexpected feature appears in the
PL. At $\lambda_0 = 933.52$ nm an additional line becomes activated as it is tuned into resonance
with the lower polariton branch $|p_-\rangle$. We can rule out the possibility that this line arises
from an unidentified charge configuration of the QD, since this additional PL peak shows a pronounced
maximum for a resonance condition with the lower polariton state rather than with the uncoupled
cavity mode. This suggests that the mechanism responsible for inducing this luminescence stems from
the excitonic component of the polariton state $|p_-\rangle$.

We argue that this PL peak is due to neutral dark exciton states, activated by a combination of
strong-coupling induced resonance between the bright-polariton and dark-exciton states, and an
efficient elastic spin-flip scattering process. As the lower polariton state $|p_-\rangle$ is red
shifted due to level repulsion, it approaches the doublet of dark excitons $|D_{a,b}\rangle$; the
large cavity-exciton coupling ensures that the $|p_-\rangle$ state has substantial excitonic
component when it reaches exact resonance with the dark-exciton transition. A sketch of this
situation is given in the insets of figure~\ref{Fig3}: here resonance between the lower polariton and
the dark excitons is achieved for a cavity position of $\lambda_{\mathrm{cav}} = 933.45\
\mathrm{nm}$. The energy splitting between bright and dark exciton lines is found to be $\delta_0 =
256\mu \mathrm{eV}$, in good agreement with values of $\delta_0$ reported elsewhere
\cite{Stevenson:PRB}. Furthermore, we have independently determined $\delta_0$ from the splitting of
the double negative charged exciton $X^{2-}$ and the single negative charged biexciton $2X^{-1}$ as
proposed in \cite{Urbaszek:PRL}. We identify those lines from their spectral locations and their
power dependencies and determine an expected splitting of 0.17 nm, in close agreement to the
measurements presented in figures \ref{Fig2} and \ref{Fig3}.

In the off-resonant excitation scheme employed in our PL measurement,  bright and dark excitons are
populated with comparable probabilities, as carriers of random spin are injected into the QD. Optical
decay of the population stored in the dark excitons however requires a spin-flip to take place prior
to the optical recombination event. The strong dependence of the dark exciton PL on the detuning from
the lower polariton branch observed here suggests that an elastic spin-flip process is involved in
the optical activation of dark exciton decay. A strong candidate for this mechanism is the hyperfine
interaction between the electron in the QD and the spins of the lattice nuclei. In this process the
electron in the QD can exchange angular momentum with the lattice nuclei with negligible energy
transfer \cite{Merkulov:PRB}. This process is very inefficient for a QD outside a cavity, as the
exchange energy $\delta_0$ has to be overcome in order to flip the electron spin. However, as
$\delta_0$ is compensated by the strong-coupling induced shift of $|p_-\rangle$, we expect a
significant increase of the hyperfine induced electron spin flip rate.

Typical hyperfine interaction strengths are on  the order of $\Omega_\mathrm{N} \simeq 0.5\
\mathrm{\mu eV}$ for QDs with $\simeq 10^5$ atoms \cite{Merkulov:PRB}. The effective optical
recombination time of dark excitons via the resonant intermediate lower polariton state $|p_-\rangle$
then is given by $\gamma_p/\Omega_\mathrm{N}^2 \simeq 110\ \mathrm{ns}$, where $\gamma_p$ is the
linewidth of the lower polariton. This timescale is relatively long compared to the $< 100\
\mathrm{ps}$ timescale of polariton recombination and the $\simeq 10\ \mathrm{ns}$ timescale of the
uncoupled exciton within the photonic bandgap. However, both bright and dark excitons are created
with equal probability in the QD, such that the populations of the two are comparable for pump powers
well below saturation. Even for different decay times this then leads to comparable PL intensities of
the cavity induced dark exciton luminescence and the PL from the polariton states, which we confirm
by multi-Lorentzian fits to our data. We note that as the pump power is increased, the dark excitons
can make transitions to other charge configurations by the capture of additional carriers, which
results in a depletion of the dark exciton population. In this case we expect a reduced strength of
the cavity mediated dark exciton PL. Indeed we have verified this from power dependent measurements
on a control device.

In conclusion we show that the strong-coupling limit of cavity QED can be used as a powerful
spectroscopic tool, providing an unambiguous identification of QD spectral features. Our findings
could prove to be useful in more complicated systems such as coupled-QDs where the multitude of
emission lines render precise identification of PL lines more difficult.

The authors would like to acknowledge Andreas Reinhard for many helpful discussions. This work is
supported by NCCR Quantum Photonics (NCCR QP), research instrument of the Swiss National Science
Foundation (SNSF).

\end{document}